\documentclass[%
 aip,groupedaddress,
 jmp,%
 amsmath,amssymb,
preprint,%
]{revtex4-1}

\usepackage{graphicx}% Include figure files
\usepackage{dcolumn}% Align table columns on decimal point
\usepackage{bm}% bold math
\def\apj{{\em The Astrophsical Journal}}

\def\apjl{{\em The Astrophsical Journal Letters}}

\def\grl{{\em Geophys. Res. Lett.}}

\def\prl{{\em Phys. Rev. Lett.}}
\def\pra{{\em Physical Review A}}

\def\nat{{\em Nature}}

\def\sci{{\em Science}}
\def\pp{{\em Plasma Physics}}
\def\pop{{\em Phys. Plasma}}
\def\pof{{\em Phys. Fluid}}

\def\physrep{{\em Physics Reports}}
\def\RMP{{\em Rev.~Mod.~Phys.}}

\begin{document}

\preprint{APS/123-QED}

\title{The Adiabatic Phase Mixing and Heating of Electrons  in Buneman Turbulence}% Force line breaks with \\
\author{H. Che\textsuperscript{1}, J. F. Drake\textsuperscript{2}, M. Swisdak\textsuperscript{2}, M. L. Goldstein\textsuperscript{1}}
%\email{hche@umd.edu}
\affiliation{1:Goddard Space Flight Center, NASA, Greenbelt, MD, 20771, USA}
%\affiliation{2: CIPS, University of Colorado, Boulder, CO, 80302, USA }
%\affiliation{2: Department of Physics, University of Maryland, College Park, MD, 20742, USA}
\affiliation{2: IREAP, University of Maryland, College Park, MD, 20742, USA}%Lines break automatically or can be forced with 
%\affiliation{3:Heliospheric Physics Laboratory, Goddard Space Flight Center, NASA, Greenbelt, MD, 20771, USA}
%\affiliation{3: IPST, University of Maryland, College Park, MD, 20742, USA}

\date{\today}% It is always \today, today
\begin{abstract}
 The nonlinear development of the strong Buneman instability and the associated fast electron heating in thin current layers with $\Omega_e/\omega_{pe} <1$ are explored. Phase mixing of the electrons in wave potential troughs and a rapid increase in temperature are observed during the saturation of the instability. We show that the motion of trapped electrons can be described using a Hamiltonian formalism in the adiabatic approximation. The process of separatrix crossing as electrons are trapped and de-trapped is irreversible and guarantees that the resulting electron energy gain is a true heating process. 
\end{abstract}

%\pacs{52.35.Vd, 52.35 Py, 52.35 Qz}% PACS, the Physics and Astronomy
                             % Classification Scheme
\maketitle
%\section{introduction}
  The exploration of how waves and particles interact in strong turbulence has been an important challenge in plasma physics. \cite{kadomtsev65book,neil65pof,dupree66pof,sagdeev69book,galeev75sov,goldman84romp,krommes02phr,yam09pop,beni12pop} Using particle-in-cell simulations we explore the nonlinear development and nonlinear wave-particle interactions of the Buneman instability to reveal how particle acceleration and heating take place.  The Buneman instability\cite{buneman58prl} is driven by the relative drift between ions and electrons. Its quasi-linear theory is well understood, but strong Buneman turbulence is still a subject with open questions though it  has been widely discussed\cite{davidson70prl,ishihaha81pof,hirose82pof,cargill88apjl}. The previous work either did not consider the trapping regime (where the wave electric field is large enough to trap thermal particles) or treated it under the assumption that the particle heating growth rate was slow compared  to the instability. We investigate the regime in which rapid electron heating takes place near the saturation of the Buneman instability when the particle's bounce rate in the wave potential is far larger than the growth rate of the instability. As a consequence, the trapped particle's motion is approximately adiabatic. Heating is thus a consequence of coherent trapping, phase mixing and de-trapping of the particles. Our simulations also demonstrate the difference between the nonlinear development of the Buneman instability and an idealized adiabatically-growing single sine wave, which supports that the heating can be achieved by adiabatic motion and de-trapping .
%These results support an nonlinear landau damping theory proposed by Neil in 1965\cite{neil65pof} and intensively studied in the recent years \cite{yam09pop,beni12pop}, which have noticed the importance of the adiabatic motion at nonlinaer stage of electrostatic wavepacket and the nonadiabatic trapping and de-trapping, but this theory is empirical, thus various models are built on different assumptions. 

Electron heating as a result of the Buneman instability is associated with the intense electron current layers formed during magnetic reconnection\cite{drake03sci,che10grl,kho10prl}, shocks\cite{cargill88apjl,riq09apj,matsumoto12apj} and turbulent energy cascades to sub-proton scales\cite{alex09prl,sah10prl}. In particular, understanding how kinetic turbulence transfers momentum and energy is important for revealing the role of anomalous resistivity in magnetic reconnection, which has been a long-standing puzzle\cite{kulsrud05pop,che11nat}.

 We propose a new mechanism that is responsible for extremely fast electron heating, in a few tens of electron plasma periods, during the nonlinear evolution of the Buneman instability. The dynamics is dominated by the coherent trapping and de-trapping of streaming electrons (with drift $v_{de}$) in the nearly non-propagating electric field from the instability. The wave amplitude grows until  nearly all of the streaming electrons have been trapped. Thus the electrostatic potential at saturation is approximately given by $e\phi\sim m_ev_{de}^2$. The bounce frequency $\omega_b=k_0\sqrt{e\phi/m_e}\sim \omega_{pe}$ of electrons trapped in the potential $\phi$ greatly exceeds the characteristic growth rate of the wave $\gamma \sim (m_e/m_i)^{1/3}\omega_{pe}$. As a result, the electrons trapped in the growing potential behave adiabatically, preserving their phase space area as the wave amplitude slowly changes in time. Phase mixing of the electrons in the wave potential troughs guarantees that, as the wave amplitude decreases following saturation, the de-trapping of electrons leaves a distribution of particles that forms a velocity-space plateau over the  interval ($-v_{de}, v_{de}$). The process of separatrix crossing as electrons are trapped and de-trapped is irreversible and guarantees that the resulting electron energy gain $\sim m_e v_{de}^2/2$ is a true heating process. 
 
%\section{Simulation Results}
We carry out 3D PIC simulations with strong electron drifts in an inhomogeneous current-carrying plasma with a guide field.  We apply no external perturbations to initiate reconnection, and consequently reconnection does not develop during relatively short duration of the simulation. We specify the initial magnetic  field to be $B_x/B_0=\tanh[(y-L_y/2)/w_0]$, where $B_0$ is the asymptotic amplitude of $B_x$, and $w_0$ and $L_y$ are the half-width of the initial current sheet and the box size in the $y$ direction, respectively. The guide field $B_z^2 = B^2-B_x^2$ is chosen so that the total field $B$ is constant. In our simulation, we take $w_0 = 0.1 d_i$ and $B=\sqrt{26} B_0$, where $d_i= c/\omega_{pi}$ and $\omega_{pi}$ is the ion plasma frequency. The initial temperature is $T_{e0} =T_{i0} =0.04 m_i c_{A0}^2$, the mass ratio is $m_i/m_e=100$, and  $\Omega_e \sim 0.625 \omega_{pe}$, where  $c_{A0}= B_0/(4 \pi n_0 m_i)^{1/2}$ is the asymptotic ion Alfv\'en wave speed.  The simulation domain has dimensions $L_x \times L_y \times L_z= 0.5 \times 1 \times 4$ $d_i$ with periodic boundaries in $x$ and $z$ and a conducting boundary in $y$. The initial electron drift along $z$, $v_{de}\sim 10 c_{A0}$, is above the threshold for triggering the Buneman instability.
\begin{figure}
\centering
\includegraphics[scale=0.8, trim=0 0 0 0,clip]{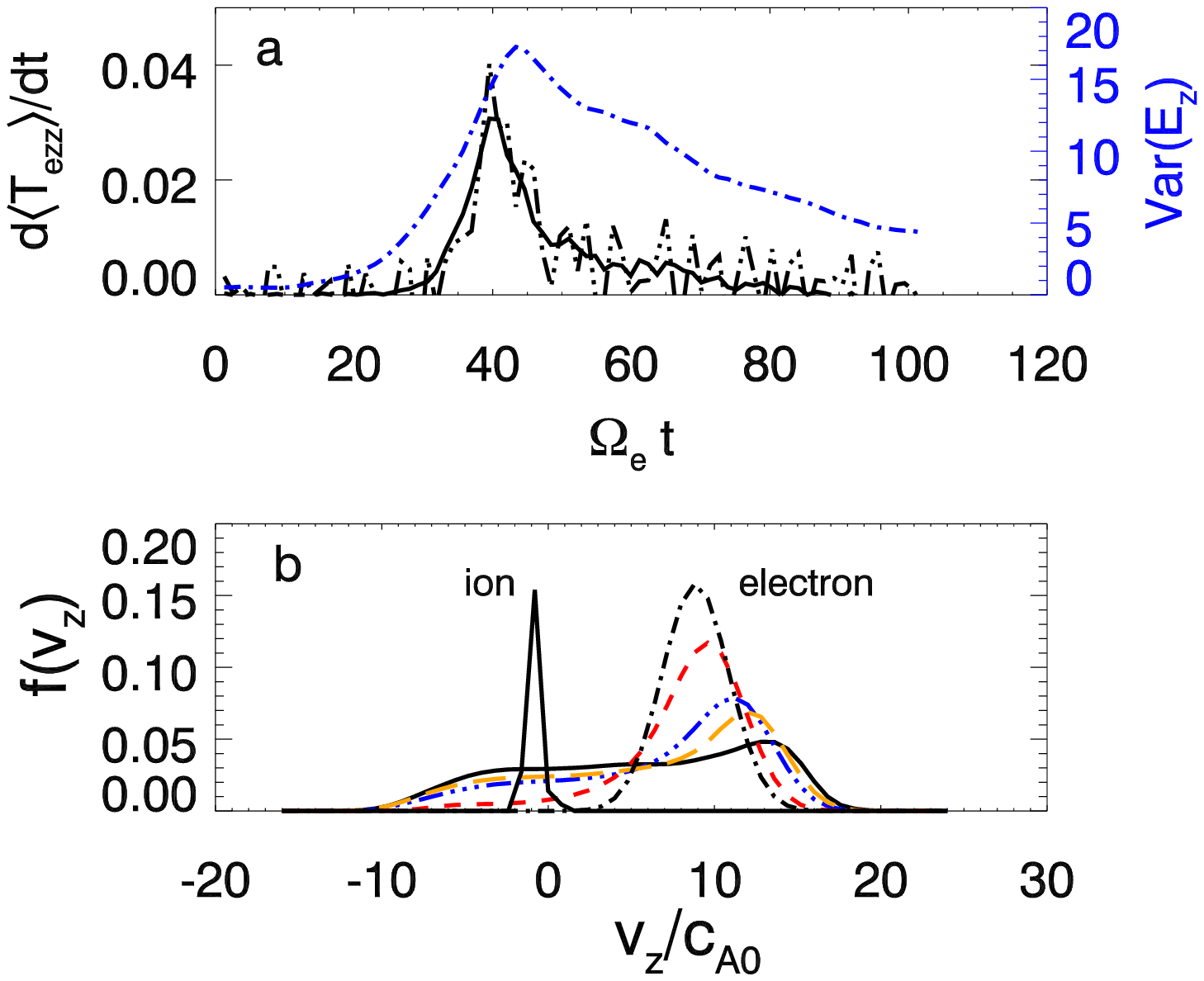} 
\includegraphics[scale=0.8, trim=0 0 0 90,clip]{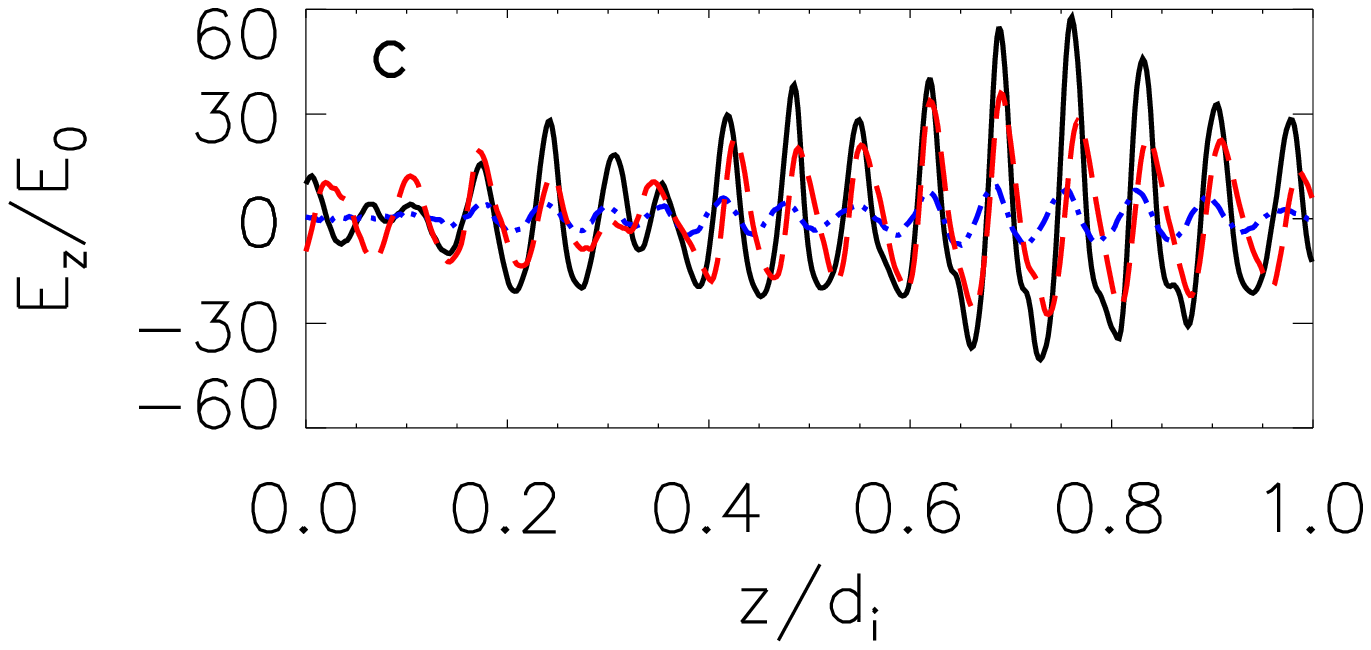} 
\caption{{\bf a}: The time evolution of $d \langle T_{ezz}\rangle/dt$ (black solid line), $\vert d W/dt\vert= -\dfrac{m_e}{e^2}d \langle  j_{ez}^2/n_e^2 \rangle /dt$ (black dot-dot-dashed line),  and the parallel electric field variance $Var(E_z)=\langle\sqrt{(E_z -\langle E_z\rangle)^2}\rangle$ (blue dot-dashed line), where $n_e$ is the electron density and $j_{ez}$ is the $z$-component of the electron current density. {\bf b}: Electron velocity  distribution functions  in the current sheet $f(v_{ez})$ at $\Omega_e t =25.5$, 38.2, 51, 63.7 and 102 plotted as black dot-dashed, red short-dashed, orange long-dashed, blue dot-dot-dashed and black solid lines respectively. The narrow ion velocity distribution function $f(v_{iz})$ at $\Omega_e t =102$ is reduced by 5 times. {\bf c}: Electric field $E_z$ parallel to $B$ at $\Omega_e t = $25.5 (blue dot-dashed), 38.2 (black solid), 63.7 (red dashed).  }
\label{eheat}
\end{figure} 

During the simulation the Buneman instability onsets around $\Omega_e t \simeq 25.5$ with a wave vector that is aligned along the magnetic field. The instability reaches its peak around $\Omega_e t \simeq 40$ and ceases around $\Omega_e t \simeq 102$, as indicated by the turbulence strength $Var(E_z)$ in Fig.\ref{eheat}a (blue dot-dashed line). The electric field $E_z$ parallel to $B_z$ abruptly increases from a few $E_0= c_{A0} B_0/c$  to $E_z\sim 40-60 E_0$ at $\Omega_e t\sim 40$ and then falls to a value $20 E_0$ at $\Omega_e t\sim 64$ (Fig.\ref{eheat}c). At the same time, the average parallel component of the electron temperature, $\langle T_{ezz}\rangle$ sharply increases, from 0.04 to 0.5, by more than a factor of 10 while the ion temperature increases only slightly. $\langle\rangle$ denotes an average over the mid-plane of the current sheet at $y=L_y/2$. The electron drift velocity decreases from $9 c_{A0}$ to $\sim 6 c_{A0}$. It is noteworthy in Fig.~\ref{eheat}a that the increase of $\langle T_{ezz}\rangle$ nearly matches the damping rate of the electron parallel kinetic energy $W$, which implies that the streaming kinetic energy is nearly fully converted into thermal energy, i.e. $\dfrac{m_e}{2}\triangle\langle v_{de}^2\rangle \approx \dfrac{1}{2}\triangle\langle T_{ezz} \rangle $, where the Boltzmann constant has been absorbed into T. Panel  b in Fig.~\ref{eheat} shows the electron velocity distribution function in the current sheet at $\Omega_e t =25.5$, 38.3, 51, 63.7 and 102. The narrow ion distribution function at $\Omega_e t =102$ is shown with a solid line. We can see that the electron velocity distribution functions $f(v_{ez})$ become flatter and broader at late times, but the significant change takes place during $\Omega_e t \sim 38 - 51$. The electron distribution functions reveal that a few electrons are accelerated to very high velocity, which is a consequence of the inductive electric field $E_z$ that maintains the integrated current. The ion velocity distribution function is slightly affected by the turbulence.
\begin{figure}
\includegraphics[scale=1, trim=50 80 0 50 ]{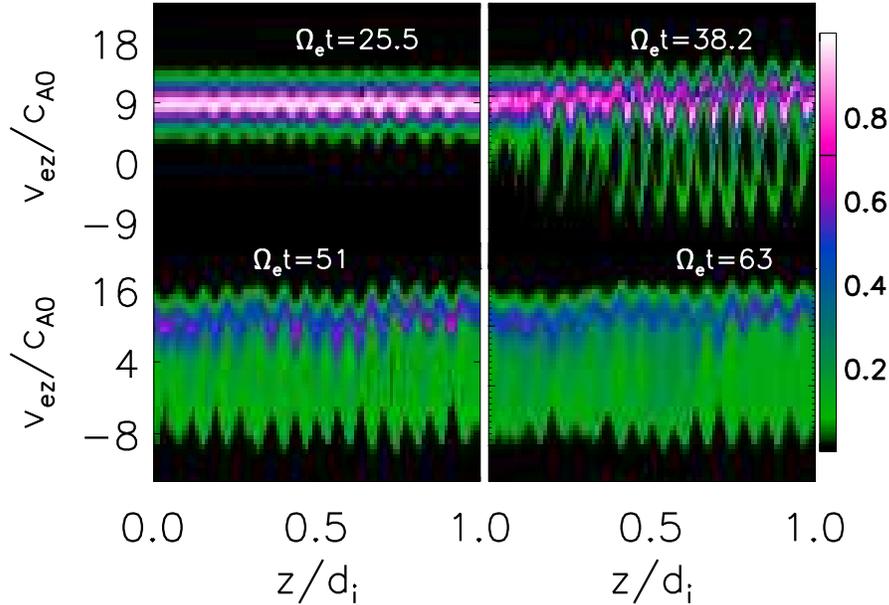} 
\caption{Electron distribution functions in phase space $(z, v_{ez})$ at $\Omega_e t = $25.5, 38.2, 51 and 63.7.}
\label{eholes}
\end{figure}

To fully understand the fast electron thermalization, we plot the electron distribution function in phase space $(z, v_{ez})$ at 
times $\Omega_e t = $ 25.5, 38.2, 51, and 63.7 in Fig.~\ref{eholes}.  We  see that at $\Omega_e t = 38.2$, electrons begin their trapped orbits in the potential wells and by $\Omega_e t = 51$ they have phase mixed along their trapped orbits. The localized electric field structure remains intact. Phase mixing does not change significantly after $\Omega_e t = 63.7$. The period of fast phase mixing is coincident with that of the flattening of electron distribution functions shown in Fig.~\ref{eheat}. The phase mixing occurs near the time of  saturation when the change in the electric field is small. We now explore the physical mechanism that produces the fast phase mixing and electron heating.

To reveal the physics behind this phase mixing and electron heating, we need to identify the properties of the instability driving the turbulence. We use a double drifting-Maxwellian kinetic model to trace the evolution of the instabilities in the simulation in the following way \cite{chephd, che09prl,che10grl}. We fit the ion and electron distribution functions averaged over $z$ at $x=0.025$ in the mid-plane $x-z$ of the current sheet at $\Omega_e t =$ 25.5, 38.2, 51, and 63.7, and then 
substitute the fits into the local dispersion relation derived from a double drifting-Maxwellian kinetic model for waves with $\Omega_{i} \ll \omega \ll \Omega_e$:
%\begin{eqnarray}
\begin{multline}
%\begin{align*}
1+\frac{2\omega_{pi}^2}{k^2 v^2_{ti}}[1+\zeta_i Z(\zeta_i)] 
+ \frac{2 \delta\omega^2_{pe}}{k^2 v^2_{zte1}}[1+I_0(\lambda) e^{-\lambda} \zeta_{e1} Z(\zeta_{e1})]\\ 
+ \frac{2 (1-\delta)\omega^2_{pe}}{k^2 v^2_{zte2}}[1+I_0(\lambda) e^{-\lambda} \zeta_{e2} Z(\zeta_{e2})]=0,
%\end{align*}
\label{bdf}
\end{multline}
%\end{equation}
%\end{eqnarray}
where  $\zeta_i=(\omega-k_{z} v_{di})/k v_{ti}$, $\zeta_{e1} = (\omega-k_{z} v_{de1})/k_{z} v_{z te1}$, $\zeta_{e2} = (\omega-k_{z} v_{de 2})/k_{z} v_{z te2}$, $\lambda=k^2_x v^2_{x te}/2\Omega^2_e$, $Z$ is the plasma dispersion function and $I_0$ is the modified Bessel function of the first kind with order zero. The thermal velocity of species $j$ is defined by  $v^2_{tj}=2T_{tj}/m_j$ and the drift speed by $v_{dj}$ which is parallel to the $z$ direction. The electron temperature takes on different values parallel and across the magnetic field while the ions are assumed to be isotropic. $\delta$ is the weight of the high velocity drifting Maxwellian. 

We numerically solve the dispersion relation in Eq.~\ref{bdf} and obtain the unstable modes at $\Omega_e t =$ 25.5, 38.2,
51, 63.7 and 102. We find that the Buneman instability dominates. The growth rate of the fastest growing Buneman mode decreases with time from $\gamma \sim 0.12 \omega_{pe}$ (close to the linear value given by Ishihara et al \cite{ishihaha81pof} $\gamma\sim \sqrt{3}/2(m_e/2m_i)^{1/3}(1-(m_e/2m_i)^{1/3}/2)\omega_{pe}\sim 0.13 \omega_{pe}$) at $\Omega_e t=25.5$ to $\gamma \sim 0.06 \omega_{pe}$ at $\Omega_e t=38.2$ and $\gamma \sim 0.006 \omega_{pe}$ at $\Omega_e t=102$. The frequency of the fastest growing mode is about $\omega_0 \sim 0.013 \omega_{pe}$ and its wavenumber $k_0 d_i$ decreases from 90 to 75. The phase speed $v_p$ increases slightly with time and has a value of $v_p\sim 0.05 c_{A0}$. A transient two stream instability with growth rate $\gamma \sim 0.006 \omega_{pe}$ appears at $\Omega_e t=51$ and is stable by $\Omega_e t=63.7$. An oblique lower hybrid instability develops with growth rate $\gamma \sim 0.02 \omega_{pe}$  after $\Omega_e t=51$.  

\begin{figure}
\includegraphics[scale=0.8, trim=0 10 0 50,clip]{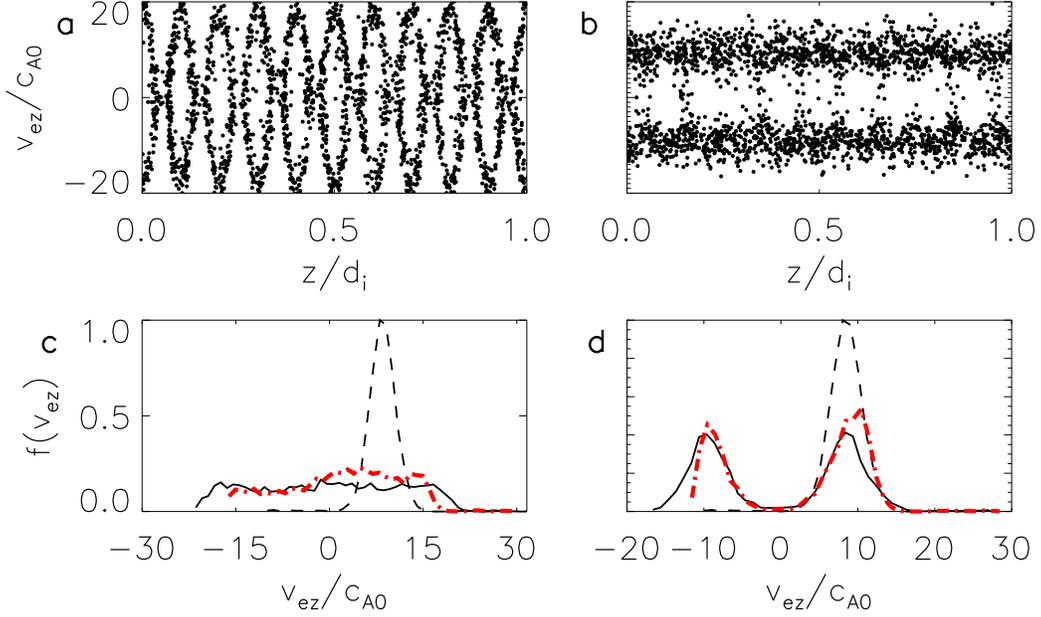} 
\caption{Panel (a) and (b) are electron distributions in phase space $z, v_{ez}$ at the middle and the end time of the test particle simulation with $E_{max}\sim 80$. The black dashed lines in Panel (c) and (d) are the initial test electron velocity distribution functions. Panel (c) is at the middle time of the test particle simulations and panel (d) is at the end time of the simulations. The black solid lines are for the test particle simulation with $E_{max}\sim 80$ and the red dot-dashed line are for the test particle simulation with $E_{max}\sim 40$.}
\label{sin}
\end{figure} 

It's interesting to notice that during  $\Omega_e t \sim 38-64$, the typical parallel electric field is about $40 E_0$ and the wavenumber of the fastest mode is $k_0 d_i \sim 90$. The corresponding bounce frequency is $\omega_b = k_0 v_b /\sqrt{2}\sim  \omega_{pe} $, where $v_b \sim \sqrt{2e\phi/m_e} \sim 10 c_{A0}$. The bounce rate is more than ten times larger than the growth rate. Thus the amplitude of the electric field  $E_z$  evolves slowly compared with $\omega_b^{-1}$ during this interval. By assuming the slowly evolving and slowly propagating (i.e. $z-v_p t \approx  z$) wave potential is $\phi(z,t)$, the electron Hamiltonian can be approximated as:
\begin{equation}
H \approx \frac{m_e}{2} v_{ez}^{ 2}-e\phi(z,t).
\label{traj1}
\end{equation}

Eq. (\ref{traj1}) shows that during $\Omega_e t =38-64$, it is possible to choose $z$ and $v_{ez}$ as two approximate Hamiltonian canonical coordinates so that the area $S=\frac{1}{2 \pi}\oint v_{ez}dz$ enclosed by the electron's trajectory in phase space $(z, v_{ez})$  is an adiabatic invariant for trapped electrons, where $v_{ez}=\sqrt{2(W(t,z)+e\phi)/m_e}$ and $W(t, z)$ is the electron's total energy. With the slow variation of the electric field, the electron's trajectory in phase space $(z, v_{ez})$ becomes narrower in $z$ and longer in $v_{ez}$ as $E_z$ increases and becomes wider in $z$ and shorter in $v_{ez}$ as $E_z$ decreases. The electrons are trapped when the electric field grows and are de-trapped when the electric field decays. The trapping and de-trapping are non-adiabatic due to the change of the phase area inside and outside of the wave potential \cite{cary86pra}.  The final electron velocity depends on whether it crosses the upper or lower separatrix as it is de-trapped. The upper (lower) separatrix crossing results in a positive (negative) velocity in the wave frame. 

To investigate how the adiabatic process converts kinetic energy into thermal energy through non-adiabatic separatrix crossings of the wave potential, we perform two test particle simulations with 5000 electrons in one single standing wave $E=E_{z0} e^{\gamma t} \sin k z$. We take $k d_i\sim 80 \sim k_0$; $E_{z0}$ and $\vert\gamma\vert$ are constant and small. They satisfy $\omega_b /\vert\gamma\vert \sim 80$ at the peak value of $E_{z}$. $\gamma> 0$ for the first half of the total duration and $\gamma< 0$ for the second half so that  $E$ grows and decays sufficiently slowly to guarantee that the motion of the trapped particles is adiabatic during the entire duration. The duration is determined by the peak value of $E_{max}$. We investigate the cases with $E_{max}\sim 40$ and $E_{max} \sim 80$. The initial electron velocity distribution is a Maxwellian with a drift $v_{de}\sim 9 c_{A0}$ and $T_e=0.04 m_i c_{A0}^2$ and the electron density is uniform in space. The value $E_{max}=40$ is similar to the  peak value of $E_z$ observed when the PIC simulations can trap electrons with velocity below $v_{de}$. $E_{max}=80$ is higher than the peak value of $E_z$ observed in the PIC simulations. The test single wave with $E_{max}=80$ can trap almost all of the electrons. The results are shown in Fig.\ref{sin}.

More and more electrons are trapped as the electric field slowly increases and the most are trapped at the peak value of $E$. The slight energy difference between two trapped electrons leads to a large separation in their phase angle around their trapped orbit since their angular velocity depends on energy. Thus, at the time of the maximum trapping the trapped electrons are nearly uniformly distributed along their trajectories as shown in panel {\bf a} and the velocity distribution of trapped electrons become flat as shown in panel {\bf c}.  As $E_z$ decreases, the electron energy gain during trapping reverses and the electrons are eventually de-trapped with the same value of $W$. The total energy $W$ is symmetric with respect to positive and negative velocity. Therefore, at the end of the simulation, due to the same probability for de-trapping at the positive and negative velocity (Fig.\ref{sin} panel {\bf b}), a dip appears near zero velocity in the velocity distribution function shown in panel {\bf d}. The red lines in panel {\bf c} and {\bf d} are for $E_{max}\sim 40$ where $E_{max}$ is not strong enough to trap all of the electrons. As a result, the distribution functions are not completely symmetric around $v_{ez}=0$.
\begin{figure}
\includegraphics[scale=0.8]{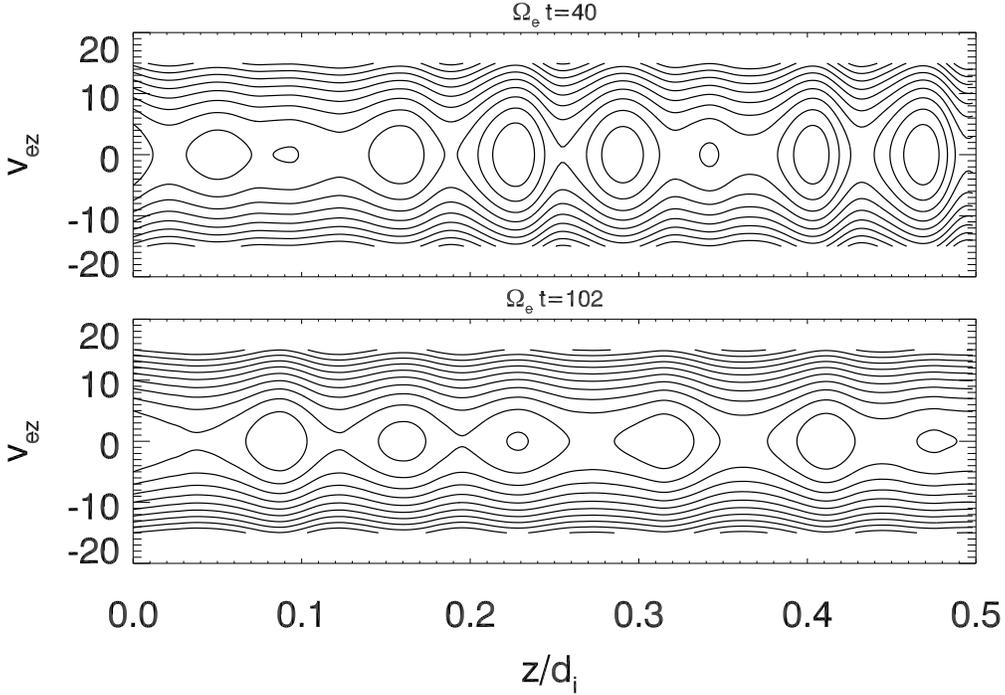} 
\caption{The contours of total electron energy in phase space $(z, v_{ez})$ at $\Omega_e t=$ 40, 102 from the PIC simulations.}
\label{island}
\end{figure} 
The red line in panel {\bf c} is similar to the electron velocity distribution function at the saturation stage of the Buneman instability  displayed in Fig.\ref{eheat}b. 

However, even at $\Omega_e t=102$ in Fig.~\ref{eheat}b, the electron velocity distribution function keeps a similar shape to that at the saturation stage rather than show a dip near zero as seen in the test particle simulation. 
There are two reasons for the missing dip in the PIC simulations. First, the wave amplitude is not spatially uniform. When the Buneman instability enters the nonlinear stage, strong wave-wave interactions cause the collapse of the uniformly distributed waves in space, and form localized solitary waves.  As a result, the trapping/de-trapping is more complex than in the sample model. In Fig.~\ref{island}, we plot the constant energy contours of electrons in phase space at $\Omega_e t =$ 40 and 102 which correspond, respectively, to the peak of the Buneman instability and late time of turbulence. We can see that the islands between $z\in [0, 0.2]$ and $z \in [0.3, 0.4]$  are longer than those between $z\in [0.2, 0.3]$ and $z \in [0.4, 0.5]$ at $\Omega_i t =40$ due to the corresponding variation of the electric field $E_z$ in $z$ as shown in Fig.~\ref{eheat}c. The long islands in phase space correspond to weak electric field and weak trapping. Second, after $\Omega_e t=51$, the electron two-stream and Buneman instabilities remain unstable, albeit weaker, and trapping continues. At the late stage of the turbulence $\Omega_e t =102$, the islands in phase space (Fig.~\ref{island}) still exist even though the islands become longer.  

In the PIC simulation, the electron trapping velocity covers the range $[-10, 10] c_{A0} \sim [-v_{de}, v_{de}] $. The heating stops when the source of kinetic energy is completely drained, i.e. the electron distribution with velocity below $v_{de}$ becomes flat, as shown in Fig. \ref{eheat} {\bf b}. Nearly half of the kinetic energy is dissipated, and the final electron temperature can be estimated by $\triangle T_{ezz}\sim \triangle W \sim m_e v_{de}^2/2\sim 0.4$  which is consistent with what is observed in the simulation (Fig. \ref{eheat}a).  

In reconnection, current sheets shrink as reconnection evolves and the Buneman instability might occur in a wider current sheet with a reduced drift but a similar growth rate $\sim (m_e/2m_i)^{1/3}\omega_{pe}$. The implications of these results in reconnection are being explored.

This research was supported by the NASA Postdoctoral Program at NASA/GSFC administered by Oak Ridge 
Associated Universities through a contract with NASA. The simulations and analysis were partially  carried out at NASA/Ames High-End Computing Capacity, at the National Energy Research Scientific  Computing Center, and at the National Institute for Computation Sciences.
%

%\newpage %Just because of unusual number of tables stacked at end
%\bibliography{plasma}% Produces the bibliography via BibTeX.

\begin{thebibliography}{26}%
\makeatletter
\providecommand \@ifxundefined [1]{%
 \@ifx{#1\undefined}
}%
\providecommand \@ifnum [1]{%
 \ifnum #1\expandafter \@firstoftwo
 \else \expandafter \@secondoftwo
 \fi
}%
\providecommand \@ifx [1]{%
 \ifx #1\expandafter \@firstoftwo
 \else \expandafter \@secondoftwo
 \fi
}%
\providecommand \natexlab [1]{#1}%
\providecommand \enquote  [1]{``#1''}%
\providecommand \bibnamefont  [1]{#1}%
\providecommand \bibfnamefont [1]{#1}%
\providecommand \citenamefont [1]{#1}%
\providecommand \href@noop [0]{\@secondoftwo}%
\providecommand \href [0]{\begingroup \@sanitize@url \@href}%
\providecommand \@href[1]{\@@startlink{#1}\@@href}%
\providecommand \@@href[1]{\endgroup#1\@@endlink}%
\providecommand \@sanitize@url [0]{\catcode `\\12\catcode `\$12\catcode
  `\&12\catcode `\#12\catcode `\^12\catcode `\_12\catcode `\%12\relax}%
\providecommand \@@startlink[1]{}%
\providecommand \@@endlink[0]{}%
\providecommand \url  [0]{\begingroup\@sanitize@url \@url }%
\providecommand \@url [1]{\endgroup\@href {#1}{\urlprefix }}%
\providecommand \urlprefix  [0]{URL }%
\providecommand \Eprint [0]{\href }%
\providecommand \doibase [0]{http://dx.doi.org/}%
\providecommand \selectlanguage [0]{\@gobble}%
\providecommand \bibinfo  [0]{\@secondoftwo}%
\providecommand \bibfield  [0]{\@secondoftwo}%
\providecommand \translation [1]{[#1]}%
\providecommand \BibitemOpen [0]{}%
\providecommand \bibitemStop [0]{}%
\providecommand \bibitemNoStop [0]{.\EOS\space}%
\providecommand \EOS [0]{\spacefactor3000\relax}%
\providecommand \BibitemShut  [1]{\csname bibitem#1\endcsname}%
\let\auto@bib@innerbib\@empty
%</preamble>
\bibitem [{\citenamefont {{Kadomtsev}}(1965)}]{kadomtsev65book}%
  \BibitemOpen
  \bibfield  {author} {\bibinfo {author} {\bibfnamefont {B.~B.}\ \bibnamefont
  {{Kadomtsev}}},\ }\href@noop {} {\emph {\bibinfo {title} {{Plasma
  turbulence}}}}\ (\bibinfo  {publisher} {New York: Academic Press, 1965},\
  \bibinfo {year} {1965})%
\bibitem [{\citenamefont {{O'Neil}}(1965)}]{neil65pof}%
  \BibitemOpen
  \bibfield  {author} {\bibinfo {author} {\bibfnamefont {T.}~\bibnamefont
  {{O'Neil}}},\ }\bibfield  {title} {\enquote {\bibinfo {title} {{Collisionless
  Damping of Nonlinear Plasma Oscillations}},}\ }\href {\doibase
  10.1063/1.1761193} {\bibfield  {journal} {\bibinfo  {journal} {\pof}\
  }\textbf {\bibinfo {volume} {8}},\ \bibinfo {pages} {2255--2262} (\bibinfo
  {year} {1965})}%
\bibitem [{\citenamefont {{Dupree}}(1966)}]{dupree66pof}%
  \BibitemOpen
  \bibfield  {author} {\bibinfo {author} {\bibfnamefont {T.~H.}\ \bibnamefont
  {{Dupree}}},\ }\bibfield  {title} {\enquote {\bibinfo {title} {{A
  Perturbation Theory for Strong Plasma Turbulence}},}\ }\href@noop {}
  {\bibfield  {journal} {\bibinfo  {journal} {\pof}\ }\textbf {\bibinfo
  {volume} {9}},\ \bibinfo {pages} {1773--1782} (\bibinfo {year}
  {1966})}%
\bibitem [{\citenamefont {{Sagdeev}}\ and\ \citenamefont
  {{Galeev}}(1969)}]{sagdeev69book}%
  \BibitemOpen
  \bibfield  {author} {\bibinfo {author} {\bibfnamefont {R.~Z.}\ \bibnamefont
  {{Sagdeev}}}\ and\ \bibinfo {author} {\bibfnamefont {A.~A.}\ \bibnamefont
  {{Galeev}}},\ }\href@noop {} {\emph {\bibinfo {title} {{Nonlinear Plasma
  Theory}}}}\ (\bibinfo  {publisher} {Nonlinear Plasma Theory, New York:
  Benjamin},\ \bibinfo {year} {1969})%
\bibitem [{\citenamefont {{Galeev}}\ \emph {et~al.}(1975)\citenamefont
  {{Galeev}}, \citenamefont {{Sagdeev}}, \citenamefont {{Shapiro}},\ and\
  \citenamefont {{Shevchenko}}}]{galeev75sov}%
  \BibitemOpen
  \bibfield  {author} {\bibinfo {author} {\bibfnamefont {A.~A.}\ \bibnamefont
  {{Galeev}}}, \bibinfo {author} {\bibfnamefont {R.~Z.}\ \bibnamefont
  {{Sagdeev}}}, \bibinfo {author} {\bibfnamefont {V.~D.}\ \bibnamefont
  {{Shapiro}}}, \ and\ \bibinfo {author} {\bibfnamefont {V.~I.}\ \bibnamefont
  {{Shevchenko}}},\ }\bibfield  {title} {\enquote {\bibinfo {title} {{Nonlinear
  effects in an inhomogeneous plasma}},}\ }\href@noop {} {\bibfield  {journal}
  {\bibinfo  {journal} {Akademiia Nauk SSSR Otdelenie Obshchei Fiziki i
  Astronomii Nauchnaia Sessiia Moscow USSR Uspekhi Fizicheskikh Nauk}\ }\textbf
  {\bibinfo {volume} {116}},\ \bibinfo {pages} {546--548} (\bibinfo {year}
  {1975})}%
\bibitem [{\citenamefont {{Goldman}}(1984)}]{goldman84romp}%
  \BibitemOpen
  \bibfield  {author} {\bibinfo {author} {\bibfnamefont {M.~V.}\ \bibnamefont
  {{Goldman}}},\ }\bibfield  {title} {\enquote {\bibinfo {title} {{Strong
  turbulence of plasma waves}},}\ }\href@noop {} {\bibfield  {journal}
  {\bibinfo  {journal} {\RMP}\ }\textbf {\bibinfo {volume} {56}},\ \bibinfo
  {pages} {709--735} (\bibinfo {year} {1984})}%
\bibitem [{\citenamefont {{Krommes}}(2002)}]{krommes02phr}%
  \BibitemOpen
  \bibfield  {author} {\bibinfo {author} {\bibfnamefont {J.~A.}\ \bibnamefont
  {{Krommes}}},\ }\bibfield  {title} {\enquote {\bibinfo {title} {{Fundamental
  statistical descriptions of plasma turbulence in magnetic fields}},}\
  }\href@noop {} {\bibfield  {journal} {\bibinfo  {journal} {\physrep}\
  }\textbf {\bibinfo {volume} {360}},\ \bibinfo {pages} {1--4} (\bibinfo {year}
  {2002})}%
\bibitem [{\citenamefont {{Yampolsky}}\ and\ \citenamefont
  {{Fisch}}(2009)}]{yam09pop}%
  \BibitemOpen
  \bibfield  {author} {\bibinfo {author} {\bibfnamefont {N.~A.}\ \bibnamefont
  {{Yampolsky}}}\ and\ \bibinfo {author} {\bibfnamefont {N.~J.}\ \bibnamefont
  {{Fisch}}},\ }\bibfield  {title} {\enquote {\bibinfo {title} {{Simplified
  model of nonlinear Landau damping}},}\ }\href {\doibase 10.1063/1.3160604}
  {\bibfield  {journal} {\bibinfo  {journal} {\pop}\ }\textbf {\bibinfo
  {volume} {16}},\ \bibinfo {pages} {072104} (\bibinfo {year}
  {2009})}%
\bibitem [{\citenamefont {{B{\'e}nisti}}, \citenamefont {{Morice}},\ and\
  \citenamefont {{Gremillet}}(2012)}]{beni12pop}%
  \BibitemOpen
  \bibfield  {author} {\bibinfo {author} {\bibfnamefont {D.}~\bibnamefont
  {{B{\'e}nisti}}}, \bibinfo {author} {\bibfnamefont {O.}~\bibnamefont
  {{Morice}}}, \ and\ \bibinfo {author} {\bibfnamefont {L.}~\bibnamefont
  {{Gremillet}}},\ }\bibfield  {title} {\enquote {\bibinfo {title} {{The
  various manifestations of collisionless dissipation in wave propagation}},}\
  }\href {\doibase 10.1063/1.4729664} {\bibfield  {journal} {\bibinfo
  {journal} {\pop}\ }\textbf {\bibinfo {volume} {19}},\ \bibinfo {pages}
  {063110} (\bibinfo {year} {2012})},\ \Eprint {http://arxiv.org/abs/1111.1391}
  {arXiv:1111.1391 [physics.plasm-ph]} %
\bibitem [{\citenamefont {{Buneman}}(1958)}]{buneman58prl}%
  \BibitemOpen
  \bibfield  {author} {\bibinfo {author} {\bibfnamefont {O.}~\bibnamefont
  {{Buneman}}},\ }\bibfield  {title} {\enquote {\bibinfo {title} {{Instability,
  Turbulence, and Conductivity in Current-Carrying Plasma}},}\ }\href {\doibase
  10.1103/PhysRevLett.1.8} {\bibfield  {journal} {\bibinfo  {journal} {\prl}\
  }\textbf {\bibinfo {volume} {1}},\ \bibinfo {pages} {8--9} (\bibinfo {year}
  {1958})}%
\bibitem [{\citenamefont {{Davidson}}\ \emph {et~al.}(1970)\citenamefont
  {{Davidson}}, \citenamefont {{Krall}}, \citenamefont {{Papadopoulos}},\ and\
  \citenamefont {{Shanny}}}]{davidson70prl}%
  \BibitemOpen
  \bibfield  {author} {\bibinfo {author} {\bibfnamefont {R.~C.}\ \bibnamefont
  {{Davidson}}}, \bibinfo {author} {\bibfnamefont {N.~A.}\ \bibnamefont
  {{Krall}}}, \bibinfo {author} {\bibfnamefont {K.}~\bibnamefont
  {{Papadopoulos}}}, \ and\ \bibinfo {author} {\bibfnamefont {R.}~\bibnamefont
  {{Shanny}}},\ }\bibfield  {title} {\enquote {\bibinfo {title} {{Electron
  Heating by Electron-Ion Beam Instabilities}},}\ }\href {\doibase
  10.1103/PhysRevLett.24.579} {\bibfield  {journal} {\bibinfo  {journal}
  {\prl}\ }\textbf {\bibinfo {volume} {24}},\ \bibinfo {pages} {579--582}
  (\bibinfo {year} {1970})}%
\bibitem [{\citenamefont {{Ishihara}}, \citenamefont {{Hirose}},\ and\
  \citenamefont {{Langdon}}(1981)}]{ishihaha81pof}%
  \BibitemOpen
  \bibfield  {author} {\bibinfo {author} {\bibfnamefont {O.}~\bibnamefont
  {{Ishihara}}}, \bibinfo {author} {\bibfnamefont {A.}~\bibnamefont
  {{Hirose}}}, \ and\ \bibinfo {author} {\bibfnamefont {A.~B.}\ \bibnamefont
  {{Langdon}}},\ }\bibfield  {title} {\enquote {\bibinfo {title} {{Nonlinear
  evolution of Buneman instability}},}\ }\href {\doibase 10.1063/1.863392}
  {\bibfield  {journal} {\bibinfo  {journal} {\pof}\ }\textbf {\bibinfo
  {volume} {24}},\ \bibinfo {pages} {452--464} (\bibinfo {year}
  {1981})}%
\bibitem [{\citenamefont {{Hirose}}, \citenamefont {{Ishihara}},\ and\
  \citenamefont {{Langdon}}(1982)}]{hirose82pof}%
  \BibitemOpen
  \bibfield  {author} {\bibinfo {author} {\bibfnamefont {A.}~\bibnamefont
  {{Hirose}}}, \bibinfo {author} {\bibfnamefont {O.}~\bibnamefont
  {{Ishihara}}}, \ and\ \bibinfo {author} {\bibfnamefont {A.~B.}\ \bibnamefont
  {{Langdon}}},\ }\bibfield  {title} {\enquote {\bibinfo {title} {{Nonlinear
  evolution of Buneman instability. II - Ion dynamics}},}\ }\href {\doibase
  10.1063/1.863807} {\bibfield  {journal} {\bibinfo  {journal} {\pof}\ }\textbf
  {\bibinfo {volume} {25}},\ \bibinfo {pages} {610--616} (\bibinfo {year}
  {1982})}%
\bibitem [{\citenamefont {{Cargill}}\ and\ \citenamefont
  {{Papadopoulos}}(1988)}]{cargill88apjl}%
  \BibitemOpen
  \bibfield  {author} {\bibinfo {author} {\bibfnamefont {P.~J.}\ \bibnamefont
  {{Cargill}}}\ and\ \bibinfo {author} {\bibfnamefont {K.}~\bibnamefont
  {{Papadopoulos}}},\ }\bibfield  {title} {\enquote {\bibinfo {title} {{A
  mechanism for strong shock electron heating in supernova remnants}},}\ }\href
  {\doibase 10.1086/185170} {\bibfield  {journal} {\bibinfo  {journal} {\apjl}\
  }\textbf {\bibinfo {volume} {329}},\ \bibinfo {pages} {L29--L32} (\bibinfo
  {year} {1988})}%
\bibitem [{\citenamefont {{Drake}}\ \emph {et~al.}(2003)\citenamefont
  {{Drake}}, \citenamefont {{Swisdak}}, \citenamefont {{Cattell}},
  \citenamefont {{Shay}}, \citenamefont {{Rogers}},\ and\ \citenamefont
  {{Zeiler}}}]{drake03sci}%
  \BibitemOpen
  \bibfield  {author} {\bibinfo {author} {\bibfnamefont {J.~F.}\ \bibnamefont
  {{Drake}}}, \bibinfo {author} {\bibfnamefont {M.}~\bibnamefont {{Swisdak}}},
  \bibinfo {author} {\bibfnamefont {C.}~\bibnamefont {{Cattell}}}, \bibinfo
  {author} {\bibfnamefont {M.~A.}\ \bibnamefont {{Shay}}}, \bibinfo {author}
  {\bibfnamefont {B.~N.}\ \bibnamefont {{Rogers}}}, \ and\ \bibinfo {author}
  {\bibfnamefont {A.}~\bibnamefont {{Zeiler}}},\ }\bibfield  {title} {\enquote
  {\bibinfo {title} {{Formation of Electron Holes and Particle Energization
  During Magnetic Reconnection}},}\ }\href {\doibase 10.1126/\sci.1080333}
  {\bibfield  {journal} {\bibinfo  {journal} {\sci}\ }\textbf {\bibinfo
  {volume} {299}},\ \bibinfo {pages} {873--877} (\bibinfo {year}
  {2003})}%
\bibitem [{\citenamefont {{Che}}\ \emph {et~al.}(2010)\citenamefont {{Che}},
  \citenamefont {{Drake}}, \citenamefont {{Swisdak}},\ and\ \citenamefont
  {{Yoon}}}]{che10grl}%
  \BibitemOpen
  \bibfield  {author} {\bibinfo {author} {\bibfnamefont {H.}~\bibnamefont
  {{Che}}}, \bibinfo {author} {\bibfnamefont {J.~F.}\ \bibnamefont {{Drake}}},
  \bibinfo {author} {\bibfnamefont {M.}~\bibnamefont {{Swisdak}}}, \ and\
  \bibinfo {author} {\bibfnamefont {P.~H.}\ \bibnamefont {{Yoon}}},\ }\bibfield
   {title} {\enquote {\bibinfo {title} {{Electron holes and heating in the
  reconnection dissipation region}},}\ }\href {\doibase 10.1029/2010GL043608}
  {\bibfield  {journal} {\bibinfo  {journal} {\grl}\ }\textbf {\bibinfo
  {volume} {37}},\ \bibinfo {pages} {11105--+} (\bibinfo {year} {2010})},\
  \Eprint {http://arxiv.org/abs/1001.3203} {arXiv:1001.3203} %
\bibitem [{\citenamefont {{Khotyaintsev}}\ \emph {et~al.}(2010)\citenamefont
  {{Khotyaintsev}}, \citenamefont {{Vaivads}}, \citenamefont {{Andr{\'e}}},
  \citenamefont {{Fujimoto}}, \citenamefont {{Retin{\`o}}},\ and\ \citenamefont
  {{Owen}}}]{kho10prl}%
  \BibitemOpen
  \bibfield  {author} {\bibinfo {author} {\bibfnamefont {Y.~V.}\ \bibnamefont
  {{Khotyaintsev}}}, \bibinfo {author} {\bibfnamefont {A.}~\bibnamefont
  {{Vaivads}}}, \bibinfo {author} {\bibfnamefont {M.}~\bibnamefont
  {{Andr{\'e}}}}, \bibinfo {author} {\bibfnamefont {M.}~\bibnamefont
  {{Fujimoto}}}, \bibinfo {author} {\bibfnamefont {A.}~\bibnamefont
  {{Retin{\`o}}}}, \ and\ \bibinfo {author} {\bibfnamefont {C.~J.}\
  \bibnamefont {{Owen}}},\ }\bibfield  {title} {\enquote {\bibinfo {title}
  {{Observations of Slow Electron Holes at a Magnetic Reconnection Site}},}\
  }\href {\doibase 10.1103/PhysRevLett.105.165002} {\bibfield  {journal}
  {\bibinfo  {journal} {Physical Review Letters}\ }\textbf {\bibinfo {volume}
  {105}},\ \bibinfo {eid} {165002} (\bibinfo {year} {2010})}%
\bibitem [{\citenamefont {{Riquelme}}\ and\ \citenamefont
  {{Spitkovsky}}(2009)}]{riq09apj}%
  \BibitemOpen
  \bibfield  {author} {\bibinfo {author} {\bibfnamefont {M.~A.}\ \bibnamefont
  {{Riquelme}}}\ and\ \bibinfo {author} {\bibfnamefont {A.}~\bibnamefont
  {{Spitkovsky}}},\ }\bibfield  {title} {\enquote {\bibinfo {title} {{Nonlinear
  Study of Bell's Cosmic Ray Current-Driven Instability}},}\ }\href {\doibase
  10.1088/0004-637X/694/1/626} {\bibfield  {journal} {\bibinfo  {journal}
  {\apj}\ }\textbf {\bibinfo {volume} {694}},\ \bibinfo {pages} {626--642}
  (\bibinfo {year} {2009})},\ \Eprint {http://arxiv.org/abs/0810.4565}
  {arXiv:0810.4565} %
\bibitem [{\citenamefont {{Matsumoto}}, \citenamefont {{Amano}},\ and\
  \citenamefont {{Hoshino}}(2012)}]{matsumoto12apj}%
  \BibitemOpen
  \bibfield  {author} {\bibinfo {author} {\bibfnamefont {Y.}~\bibnamefont
  {{Matsumoto}}}, \bibinfo {author} {\bibfnamefont {T.}~\bibnamefont
  {{Amano}}}, \ and\ \bibinfo {author} {\bibfnamefont {M.}~\bibnamefont
  {{Hoshino}}},\ }\bibfield  {title} {\enquote {\bibinfo {title} {{Electron
  Accelerations at High Mach Number Shocks: Two-dimensional Particle-in-cell
  Simulations in Various Parameter Regimes}},}\ }\href {\doibase
  10.1088/0004-637X/755/2/109} {\bibfield  {journal} {\bibinfo  {journal}
  {\apj}\ }\textbf {\bibinfo {volume} {755}},\ \bibinfo {eid} {109} (\bibinfo
  {year} {2012})},\ \Eprint {http://arxiv.org/abs/1204.6312} {arXiv:1204.6312
  [astro-ph.HE]} %
\bibitem [{\citenamefont {{Alexandrova}}\ \emph {et~al.}(2009)\citenamefont
  {{Alexandrova}}, \citenamefont {{Saur}}, \citenamefont {{Lacombe}},
  \citenamefont {{Mangeney}}, \citenamefont {{Mitchell}}, \citenamefont
  {{Schwartz}},\ and\ \citenamefont {{Robert}}}]{alex09prl}%
  \BibitemOpen
  \bibfield  {author} {\bibinfo {author} {\bibfnamefont {O.}~\bibnamefont
  {{Alexandrova}}}, \bibinfo {author} {\bibfnamefont {J.}~\bibnamefont
  {{Saur}}}, \bibinfo {author} {\bibfnamefont {C.}~\bibnamefont {{Lacombe}}},
  \bibinfo {author} {\bibfnamefont {A.}~\bibnamefont {{Mangeney}}}, \bibinfo
  {author} {\bibfnamefont {J.}~\bibnamefont {{Mitchell}}}, \bibinfo {author}
  {\bibfnamefont {S.~J.}\ \bibnamefont {{Schwartz}}}, \ and\ \bibinfo {author}
  {\bibfnamefont {P.}~\bibnamefont {{Robert}}},\ }\bibfield  {title} {\enquote
  {\bibinfo {title} {{Universality of Solar-Wind Turbulent Spectrum from MHD to
  Electron Scales}},}\ }\href {\doibase 10.1103/PhysRevLett.103.165003}
  {\bibfield  {journal} {\bibinfo  {journal} {\prl}\ }\textbf {\bibinfo
  {volume} {103}},\ \bibinfo {eid} {165003} (\bibinfo {year} {2009})},\ \Eprint
  {http://arxiv.org/abs/0906.3236} {arXiv:0906.3236 [physics.plasm-ph]}
  %
\bibitem [{\citenamefont {{Sahraoui}}\ \emph {et~al.}(2010)\citenamefont
  {{Sahraoui}}, \citenamefont {{Goldstein}}, \citenamefont {{Belmont}},
  \citenamefont {{Canu}},\ and\ \citenamefont {{Rezeau}}}]{sah10prl}%
  \BibitemOpen
  \bibfield  {author} {\bibinfo {author} {\bibfnamefont {F.}~\bibnamefont
  {{Sahraoui}}}, \bibinfo {author} {\bibfnamefont {M.~L.}\ \bibnamefont
  {{Goldstein}}}, \bibinfo {author} {\bibfnamefont {G.}~\bibnamefont
  {{Belmont}}}, \bibinfo {author} {\bibfnamefont {P.}~\bibnamefont {{Canu}}}, \
  and\ \bibinfo {author} {\bibfnamefont {L.}~\bibnamefont {{Rezeau}}},\
  }\bibfield  {title} {\enquote {\bibinfo {title} {{Three Dimensional
  Anisotropic k Spectra of Turbulence at Subproton Scales in the Solar
  Wind}},}\ }\href {\doibase 10.1103/PhysRevLett.105.131101} {\bibfield
  {journal} {\bibinfo  {journal} {\prl}\ }\textbf {\bibinfo {volume} {105}},\
  \bibinfo {eid} {131101} (\bibinfo {year} {2010})}%
\bibitem [{\citenamefont {{Kulsrud}}\ \emph {et~al.}(2005)\citenamefont
  {{Kulsrud}}, \citenamefont {{Ji}}, \citenamefont {{Fox}},\ and\ \citenamefont
  {{Yamada}}}]{kulsrud05pop}%
  \BibitemOpen
  \bibfield  {author} {\bibinfo {author} {\bibfnamefont {R.}~\bibnamefont
  {{Kulsrud}}}, \bibinfo {author} {\bibfnamefont {H.}~\bibnamefont {{Ji}}},
  \bibinfo {author} {\bibfnamefont {W.}~\bibnamefont {{Fox}}}, \ and\ \bibinfo
  {author} {\bibfnamefont {M.}~\bibnamefont {{Yamada}}},\ }\bibfield  {title}
  {\enquote {\bibinfo {title} {{An electromagnetic drift instability in the
  magnetic reconnection experiment and its importance to magnetic }},}\ }\href
  {\doibase 10.1063/1.1949225} {\bibfield  {journal} {\bibinfo  {journal}
  {\pp}\ }\textbf {\bibinfo {volume} {12}},\ \bibinfo {pages} {082301}
  (\bibinfo {year} {2005})}%
\bibitem [{\citenamefont {{Che}}, \citenamefont {{Drake}},\ and\ \citenamefont
  {{Swisdak}}(2011)}]{che11nat}%
  \BibitemOpen
  \bibfield  {author} {\bibinfo {author} {\bibfnamefont {H.}~\bibnamefont
  {{Che}}}, \bibinfo {author} {\bibfnamefont {J.~F.}\ \bibnamefont {{Drake}}},
  \ and\ \bibinfo {author} {\bibfnamefont {M.}~\bibnamefont {{Swisdak}}},\
  }\bibfield  {title} {\enquote {\bibinfo {title} {{A current filamentation
  mechanism for breaking magnetic field lines during reconnection}},}\ }\href
  {\doibase 10.1038/nature10091} {\bibfield  {journal} {\bibinfo  {journal}
  {\nat}\ }\textbf {\bibinfo {volume} {474}},\ \bibinfo {pages} {184} (\bibinfo
  {year} {2011})}%
\bibitem [{\citenamefont {{Che}}(2009)}]{chephd}%
  \BibitemOpen
  \bibfield  {author} {\bibinfo {author} {\bibfnamefont {H.}~\bibnamefont
  {{Che}}},\ }\emph {\bibinfo {title} {{Non-linear development of streaming
  instabilities in magnetic reconnection with a strong guide field}}},\
  \href@noop {} {Ph.D. thesis},\ \bibinfo  {school} {University of Maryland,
  College Park} (\bibinfo {year} {2009})%
\bibitem [{\citenamefont {{Che}}\ \emph {et~al.}(2009)\citenamefont {{Che}},
  \citenamefont {{Drake}}, \citenamefont {{Swisdak}},\ and\ \citenamefont
  {{Yoon}}}]{che09prl}%
  \BibitemOpen
  \bibfield  {author} {\bibinfo {author} {\bibfnamefont {H.}~\bibnamefont
  {{Che}}}, \bibinfo {author} {\bibfnamefont {J.~F.}\ \bibnamefont {{Drake}}},
  \bibinfo {author} {\bibfnamefont {M.}~\bibnamefont {{Swisdak}}}, \ and\
  \bibinfo {author} {\bibfnamefont {P.~H.}\ \bibnamefont {{Yoon}}},\ }\bibfield
   {title} {\enquote {\bibinfo {title} {{Nonlinear Development of Streaming
  Instabilities in Strongly Magnetized Plasma}},}\ }\href {\doibase
  10.1103/PhysRevLett.102.145004} {\bibfield  {journal} {\bibinfo  {journal}
  {\prl}\ }\textbf {\bibinfo {volume} {102}},\ \bibinfo {pages} {145004--+}
  (\bibinfo {year} {2009})},\ \Eprint {http://arxiv.org/abs/0903.1311}
  {arXiv:0903.1311 [physics.space-ph]} %
\bibitem [{\citenamefont {{Cary}}, \citenamefont {{Escande}},\ and\
  \citenamefont {{Tennyson}}(1986)}]{cary86pra}%
  \BibitemOpen
  \bibfield  {author} {\bibinfo {author} {\bibfnamefont {J.~R.}\ \bibnamefont
  {{Cary}}}, \bibinfo {author} {\bibfnamefont {D.~F.}\ \bibnamefont
  {{Escande}}}, \ and\ \bibinfo {author} {\bibfnamefont {J.~L.}\ \bibnamefont
  {{Tennyson}}},\ }\bibfield  {title} {\enquote {\bibinfo {title}
  {{Adiabatic-invariant change due to separatrix crossing}},}\ }\href {\doibase
  10.1103/PhysRevA.34.4256} {\bibfield  {journal} {\bibinfo  {journal} {\pra}\
  }\textbf {\bibinfo {volume} {34}},\ \bibinfo {pages} {4256--4275} (\bibinfo
  {year} {1986})}%
\end{thebibliography}
%


\end{document}